\documentclass[11pt]{article}
\usepackage{bbm}
\usepackage[final]{graphics}
\usepackage{amsmath}
\usepackage{amsfonts,amsbsy}
\usepackage{amssymb}
\usepackage{bm}
\usepackage{color}

\def\empile#1\over#2{\mathrel{\mathop{\kern 0pt#1}\limits_{#2}}}
\def\bs{\boldsymbol}
\def\wt#1{\widetilde{#1}}

\def\TODO#1{}

\def\p{{\boldsymbol p}}

\def\x{{\boldsymbol x}}
\def\y{{\boldsymbol y}}

\def\u{{\boldsymbol u}}
\def\v{{\boldsymbol v}}

\newcommand{\slL}{\raise.15ex\hbox{$/$}\kern-.53em\hbox{$L$}}
\newcommand{\slP}{\raise.15ex\hbox{$/$}\kern-.53em\hbox{$P$}}
\newcommand{\slD}{\raise.15ex\hbox{$/$}\kern-.67em\hbox{$D$}}
\newcommand{\slp}{\raise.1ex\hbox{$/$}\kern-.63em\hbox{$p$}}
\newcommand{\slq}{\raise.1ex\hbox{$/$}\kern-.53em\hbox{$q$}}
\newcommand{\slv}{\raise.1ex\hbox{$/$}\kern-.63em\hbox{$v$}}
\newcommand{\slR}{\raise.15ex\hbox{$/$}\kern-.53em\hbox{$R$}}
\newcommand{\slQ}{\raise.15ex\hbox{$/$}\kern-.53em\hbox{$Q$}}
\newcommand{\slK}{\raise.15ex\hbox{$/$}\kern-.53em\hbox{$K$}}
\newcommand{\slk}{\raise.15ex\hbox{$/$}\kern-.53em\hbox{$k$}}
\newcommand{\slSigma}{\raise.15ex\hbox{$/$}\kern-.53em\hbox{$\Sigma$}}
\newcommand{\slcalP}{\raise.15ex\hbox{$/$}\kern-.63em\hbox{$\cal P$}}
\newcommand{\slcalA}{\raise.15ex\hbox{$/$}\kern-.63em\hbox{$\cal A$}}
\newcommand{\slA}{\raise.15ex\hbox{$/$}\kern-.73em\hbox{$A$}}
\newcommand{\slbfA}{\raise.15ex\hbox{$/$}\kern-.73em\hbox{${\imb A}$}}
\newcommand{\slpartial}{\raise.15ex\hbox{$/$}\kern-.53em\hbox{$\partial$}}
\newcommand{\sla}{\raise.15ex\hbox{$/$}\kern-.53em\hbox{$a$}}
\newcommand{\slb}{\raise.15ex\hbox{$/$}\kern-.53em\hbox{$b$}}
\newcommand{\slc}{\raise.15ex\hbox{$/$}\kern-.53em\hbox{$c$}}
\newcommand{\slC}{\raise.15ex\hbox{$/$}\kern-.63em\hbox{$C$}}
\newcommand{\sln}{\raise.15ex\hbox{$/$}\kern-.575em\hbox{$n$}}

\begin{document}

\date{}

\title{\bf Statistical fluctuations of correlators\\ in the Color Glass Condensate}
\author{Fran\c cois Gelis${}^{~a}$, Naoto Tanji${}^{~b}$}
\maketitle
 \begin{center}
   \begin{itemize}
  \item[{\bf a.}] Institut de Physique Th\'eorique, Universit\'e Paris-Saclay\\
 CEA, CNRS, F-91191 Gif-sur-Yvette, France
  \item[{\bf b.}] European Centre for Theoretical Studies in Nuclear Physics\\ and Related Areas (ECT*) 
  and Fondazione Bruno Kessler\\
Strada delle Tabarelle 286, I-38123 Villazzano, Italy
  \end{itemize}
 \end{center}
\vglue 1cm

\begin{abstract}
  In the McLerran-Venugopalan model, correlators of Wilson lines are
  given by an average over a Gaussian ensemble of random color
  sources. In numerical implementations, these averages are
  approximated by a Monte-Carlo sampling. In this paper, we study the
  statistical error made with such a sampling, with emphasis on the
  momentum dependence of this error. Using the example of the dipole
  amplitude, we consider various approximants that are all equivalent
  in the limit of infinite statistics but differ with finite
  statistics and compare their statistical errors. For correlation
  functions that are translation invariant, we show that averaging
  over the barycenter coordinate drastically reduces the statistical
  error and more importantly modifies its momentum dependence.
\end{abstract}
\vglue 1cm

\section{Introduction}
In the Color Glass Condensate (CGC) effective theory 
\cite{Iancu:2002xk,Weigert:2005us,Gelis:2010nm,Gelis:2012ri}, the degrees of freedom
internal to a high energy hadron or nucleus are split into two
categories depending on their longitudinal momentum in the observer's
frame. The dynamical evolution of the fast modes is slowed down by
time dilation, and therefore they are approximated as static
constituents that carry a color current. The slow modes cannot be
approximated in this way. Instead, they are treated as usual gauge
fields, eikonally coupled to the color current produced by the fast
modes. In this approximation, the evaluation of scattering amplitudes
involving one or more high energy hadrons in the initial state can be
formulated in terms of a Yang-Mills theory coupled to an external
color current. However, it is important to realize that this current
has event-by-event fluctuations. Indeed, this current reflects the
precise configuration of the fast partons of the projectile at the
time of the collision, which of course differs from one event to the
next one.

A simple and popular model for the fluctuations of this color current
is the McLerran-Venugopalan (MV) model \cite{McLerran:1993ni,McLerran:1993ka}, which states that for a large
enough projectile the fluctuations are Gaussian (one may view this as
a consequence of the central limit theorem). More precisely, in the MV
model, the color current of a high energy projectile moving in the
$+z$ direction is  parameterized as follows,
\begin{align}
  &J^\mu_a(x)=\delta^{\mu +} \rho_a(x^-,\x_\perp),\nonumber\\
  &\big<\rho_a(x^-,\x_\perp)\big>=0,\nonumber\\
  &  \big<\rho_a(x^-,\x_\perp)\rho_b(y^-,\y_\perp)\big>=\mu^2(x^-)\,\delta_{ab}\delta(x^--y^-)\delta(\x_\perp-\y_\perp)\; .
    \label{eq:MVsources}
\end{align}
where $\rho_a$ is a function that describes the space-time
distribution of the color charges that carry this current. For a
projectile moving in the $+z$ direction, as is the case in this
example, this distribution is independent of $x^+$ thanks to the large
Lorentz boost factor affecting the projectile: in the frame of an
observer, the internal dynamics of the projectile appears completely
frozen --and thus $x^+$ independent-- on the timescales of the
collision itself. The second equation states that the color charge
density is zero on average, and the third one defines the variance of
its (Gaussian) fluctuations. The MV model further assumes that the
correlations among the color charges are local both in $x^-$ and
$\x_\perp$. Although it is now understood that these correlations
evolve with the energy of the projectile according to the JIMWLK
equation \cite{JalilianMarian:1997jx,JalilianMarian:1997gr,Iancu:2000hn,Ferreiro:2001qy}, 
leading to non-Gaussiannities at higher energies, Gaussian
models such as the MV model remain at the center of most of the CGC
phenomenology.

In this framework, any observable may be viewed as a functional of the
source $\rho_a$, and its expectation value is obtained by a Gaussian
average based on eq.~(\ref{eq:MVsources}).  For simple observables,
this Gaussian average may be performed analytically. However, there
are many situations where the dependence of the observable on $\rho_a$
can only be obtained numerically (for instance, the inclusive spectrum
of gluons produced in nucleus-nucleus collisions depends on solutions
of the classical Yang-Mills equations in the strong field regime, that
are not known analytically \cite{Krasnitz:2000gz,Krasnitz:2001qu,Lappi:2003bi}), 
which implies that the average over the
color sources must be performed numerically as well, by a Monte-Carlo
sampling of the ensemble of sources. In this case, it is important to
estimate the statistical errors made in this sampling (assuming that
the algorithm that gives the observable in terms of $\rho_a$ has
negligible errors). More specifically, one would like to know
\begin{itemize}
\item How do the statistical errors depend on the number of samples
  used in the average?
\item How do the statistical errors depend on momentum? In particular,
  is the relative error uniform over the entire parameter space?
\end{itemize}
The goal of this paper is to address these questions, by using the
dipole amplitude (defined below) as an example. Although we use this
explicit example as the support of our discussion, our results
concerning the statistical errors in the MV model remain qualitatively
true in general.

Our paper is organized as follows. In the section \ref{sec:setup}, we
define the dipole amplitude, both in the continuum and in a lattice
discretization. In the section \ref{sec:MC}, we consider several
``naive'' ways of evaluating this observable by Monte-Carlo and study
the associated statistical errors. In the section \ref{sec:self}, we
consider an improvement that exploits the translation invariance in
the transverse plane, consisting in integrating over the barycenter of
the transverse coordinates. We show that this modification leads to
much better numerical results, and we explain analytically why this
change modifies the momentum dependence of the statistical
error. Finally, the section \ref{sec:concl} contains a summary and
conclusions.

\section{Setup of a simple example}
\label{sec:setup}
\subsection{Continuous formulation}
The dipole amplitude is the simplest observable that one encounters in
the CGC framework. It appears for instance in the
scattering cross-section of a quark off a dense nuclear target
\cite{Gelis:2001da,Gelis:2002ki}. In coordinate space, it is a two-point
function made with the trace of two Wilson lines in the light-cone
direction,
\begin{align}
  &C(\x_\perp,\y_\perp)\equiv \frac{1}{N_c}\,\Big<{\rm tr}\,\left(U(\x_\perp)U^\dagger(\y_\perp)\right)\Big>,\nonumber\\
  &U(\x_\perp)\equiv{\rm P}\,
    \exp\Big\{ig\int_{-\infty}^{+\infty}dx^-\;A_a^+(x^-,\x_\perp)t^a\Big\},
    \label{eq:Ccoord-def}
\end{align}
where $g$ is the strong coupling constant, ${\rm P}$ denotes an
operator ordering in the $x^-$ direction and $A_a^+$ is the $+$
component of the target color field in Lorenz gauge.  The latter is
related to the configuration $\rho_a$ of the color charges in the
nucleus by
\begin{eqnarray}
  -{\bs\nabla}_\perp^2\,A_a^+(x^-,\x_\perp)=\rho_a(x^-,\x_\perp).
  \label{eq:YM}
\end{eqnarray}
In eqs.~(\ref{eq:Ccoord-def}), the brackets $\big<\cdots\big>$ denote
a Gaussian average over the $\rho_a$'s.  Note that if we assume that
the quantity $\mu^2(x^-)$ in eq.~(\ref{eq:MVsources}) does not depend
on the transverse position, then the system is invariant under
translations in the transverse plane, and $C(\x_\perp,\y_\perp)$
depends only on the difference of coordinates $\x_\perp-\y_\perp$.
The quantity that we shall discuss mostly in this paper is the Fourier
transform of the $2$-point function $C(\x_\perp,\y_\perp)$ with
respect to $\x_\perp-\y_\perp$,
\begin{align}
C(\p_\perp)\equiv \int d^2\x_\perp\;e^{i\p_\perp\cdot\x_\perp}\,C(\x_\perp,0).
  \label{eq:Cpdef}
\end{align}
In this definition, we have fixed the second coordinate to be at the
origin of the transverse plane, but we could also use translation
invariance and write this as
\begin{align}
  C(\p_\perp)\equiv \int \frac{d^2\x_\perp d^2\y_\perp}{{\cal S}_\perp}\;e^{i\p_\perp\cdot(\x_\perp-\y_\perp)}\,C(\x_\perp,\y_\perp),
  \label{eq:Cpdef-1}
\end{align}
where ${\cal S}_\perp$ is the (assumed large) transverse area of the
target. The first definition is closer to what one would do in a
situation without translation invariance in the transverse plane --
indeed, in that case, the Fourier transform with respect to the
coordinate difference $\x_\perp-\y_\perp$ yields a result that still
depends on the barycenter coordinate $(\x_\perp+\y_\perp)/2$. In
contrast, the second definition also averages over the barycenter
coordinate in addition to the Fourier transform.

The Gaussian average over the ensemble of
sources described in eq.~(\ref{eq:MVsources}) %is straightforward 
can be performed analytically \cite{Gelis:2001da} 
and leads to 
\begin{equation}
      \frac{1}{N_c}\, \Big<{\rm tr}\,\left(U(\x_\perp)U^\dagger(0_\perp)\right)\Big>
  =
  \exp\Big\{- C_{_F}Q_s^2\int\frac{d^2\p_\perp}{(2\pi)^2}\;\frac{1-e^{i\p_\perp\cdot\x_\perp}}{p_\perp^4}\Big\},
  \label{eq:Cx}
\end{equation}
where $C_{_F}\equiv \tfrac{N_c}{2}-\tfrac{1}{2N_c}$ is the quadratic
Casimir operator in the fundamental representation and $Q_s^2$ denotes
\begin{equation}
Q_s^2\equiv g^2\int dx^-\;\mu^2(x^-).
\end{equation}
In the right hand side of eq.~(\ref{eq:Cx}), the integral over
$\p_\perp$ has an infrared divergence, that originates from the
logarithmic long distance behavior of solutions of the Poisson
equation in two dimensions. Since the matters discussed in this paper
are unrelated with this issue, we simply regularize this divergence by
introducing a small mass in the transverse Laplacian of
eq.~(\ref{eq:YM}), $-{\bs\nabla}_\perp^2\to -{\bs\nabla}_\perp^2+m^2$,
which amounts to replacing the denominator $p^4$ by
$(p_\perp^2+m^2)^2$ in eq.~(\ref{eq:Cx}). Another important property
of eq.~(\ref{eq:Cx}) is that the resulting correlation function is
invariant when we change $\x_\perp\to-\x_\perp$. Therefore,
its Fourier transform is real. From  eq.~(\ref{eq:Cx}), it is also easy to derive the following large momentum behavior for 
$C(\p_\perp)$,
\begin{equation}
  C(\p_\perp)\approx \frac{C_{_F} Q_s^2}{p_\perp^4}+{\cal O}\left(p_\perp^{-6}\right).
  \label{eq:asympt}
\end{equation}

\subsection{Lattice formulation}
In most numerical implementations of CGC calculations, it is necessary
to discretize the transverse plane on a lattice. On a $N_\perp\times N_\perp$
square lattice with spacing $a$ between neighboring sites, we write
\begin{align}
  x=ia, \quad y=ja\quad (0\le i,j< N_\perp).
\end{align}
A function $F(\x_\perp)$ of the transverse coordinates is then
represented by a set of number $F_{ij}$, and its Fourier transform
$\wt{F}(\p_\perp)$ becomes a discrete Fourier transform,
\begin{align}
\wt{F}_{kl}\equiv a^2 \sum_{i,j} F_{ij}\,e^{2i\pi\tfrac{ik+jl}{N_\perp}},
\end{align}
and the reverse transform reads
\begin{align}
{F}_{ij}\equiv \frac{1}{(a N_\perp)^2}\sum_{k,l} \wt{F}_{kl}\,e^{-2i\pi\tfrac{ik+jl}{N_\perp}}.
\end{align}
The correspondence between the continuum momentum $p_\perp$ and the
discrete labels $k,l$ is given by the following formula
\begin{align}
  p_\perp^2\Big|_{\rm lattice}= \frac{2}{a^2}\Big(2-\cos\big(\tfrac{2\pi k}{N_\perp}\big)-\cos\big(\tfrac{2\pi l}{N_\perp}\big)\Big)
  =\frac{4}{a^2}\Big(\sin^2\big(\tfrac{\pi k}{N_\perp}\big)+\sin^2\big(\tfrac{\pi l}{N_\perp}\big)\Big).
  \label{eq:lattice-norm}
\end{align}
(This is obtained from the eigenvalues of the discrete Laplacian.)

\begin{figure}[htbp]
  \begin{center}
    \resizebox*{9cm}{!}{\includegraphics{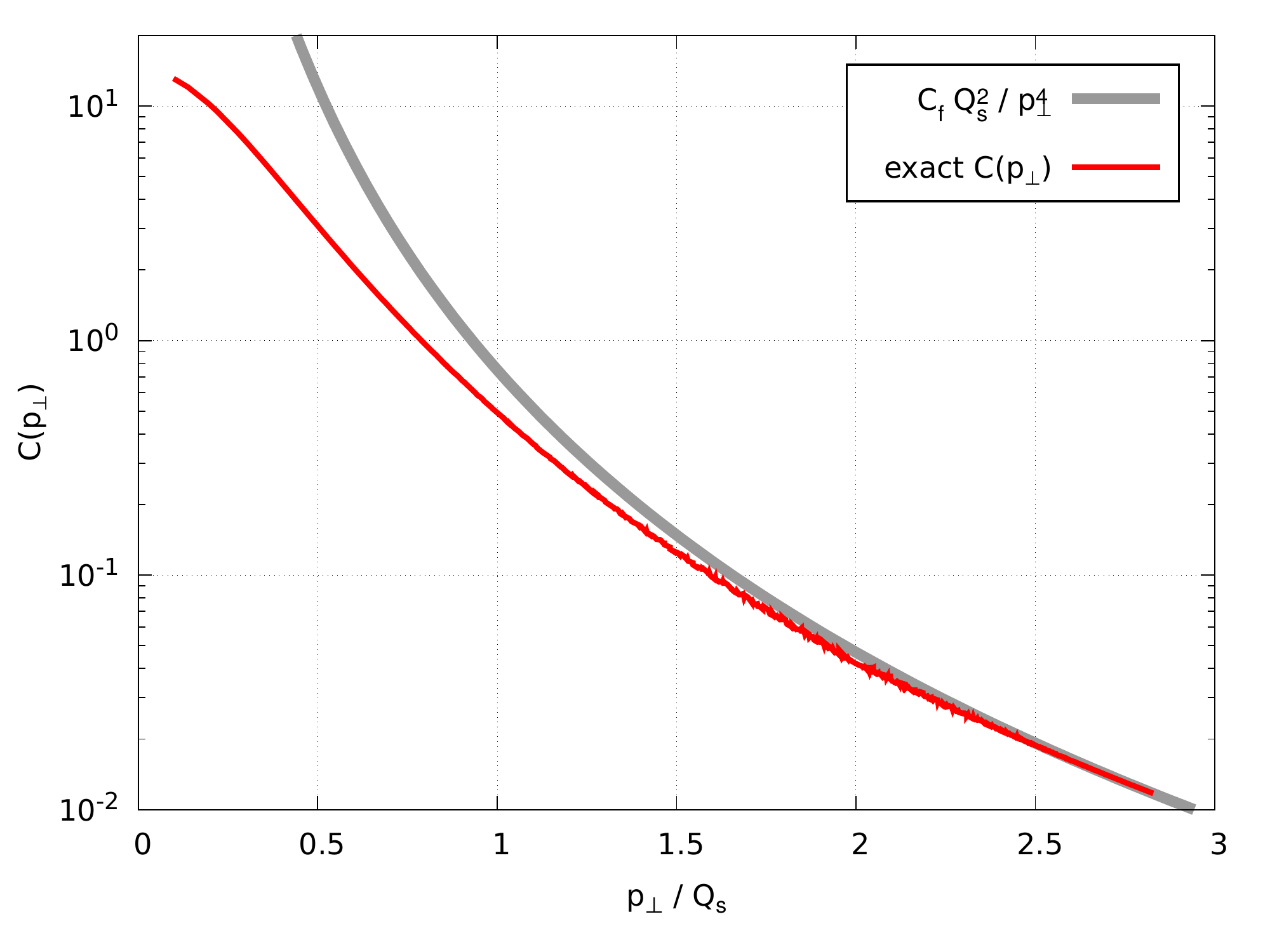}}
  \end{center}
  \caption{\label{fig:Ck}Plot of the function $C(p_\perp)$. Thin red
    line: based on a discrete version of eq.~(\ref{eq:Cx}), on a
    $64\times 64\times 20$ lattice. The mass used in the infrared
    regularization is $m^2=0.2\, Q_s^2 $. Thick gray band: asymptotic
    behavior $C_{_F} Q_s^2 /p_\perp^4$.}
\end{figure}

In the Wilson lines, the support in $x^-$ of the target field
$A^+_a(x^-,\x_\perp)$ is a small interval $[0,x^-_{\rm max}]$ where
$x^-_{\rm max}$ is inversely proportional to the collision energy.  In
order to account for the path ordering of the Wilson line, this
longitudinal interval must also be discretized
\cite{Fukushima:2007ki}. This amounts to approximating Wilson lines by
a product of ordinary exponentials.  Indeed, when the elementary
intervals in the $x^-$ direction are small, we may disregard the
non-commutativity of the $SU(N_c)$ elements within a slice and replace
the path ordered exponential for that slice by an ordinary exponential
(with the Baker-Campbell-Haussdorf formula, one may show that the
commutators produce terms of higher order in the slice thickness
$\epsilon$). With $L$ intervals of length $\epsilon$ (such that
$L\epsilon=x^-_{\rm max}$), the discrete representation of a Wilson
line is
\begin{align}
  U(\x_\perp)
  \empile{=}\over{L\to \infty}
  e^{ig\epsilon A_a^+(L\epsilon,\x_\perp) t^a }\,e^{ig\epsilon A_a^+((L-1)\epsilon,\x_\perp) t^a}\cdots e^{ig\epsilon A_a^+(\epsilon,\x_\perp) t^a},
\end{align}
where $A_a^+(l\epsilon,\x_\perp)$ is the solution of the discretized
Poisson equation (\ref{eq:YM}) in the slice $l$. In the rest of this
paper, we use a square transverse lattice of size $N_\perp=64$ 
with spacing $Q_s a=1$,
and $L=20$ slices in the $x^-$ direction. 
Furthermore, we consider $N_c=2$ just for simplicity of numerical calculations. 
In the figure \ref{fig:Ck}, we
show the result of evaluating the Fourier transform of the dipole
amplitude in this setup, starting from a discrete version of
eq.~(\ref{eq:Ccoord-def}) (this result will later be referred to as
``exact'', since the ensemble average is performed analytically), and
we compare it to the asymptotic form (\ref{eq:asympt}).

\section{Monte-Carlo evaluation without barycenter averaging}
\label{sec:MC}
For the sake of the discussion, let us now assume that we do not know
the analytical expression of eq.~(\ref{eq:Cx}). Instead, we wish to
obtain the 2-point function by a Monte-Carlo sampling of the Gaussian
distribution of the sources $\rho_a$.  In this section, we first
consider numerical estimates of the dipole amplitude that evaluate the
two-point function with one coordinate fixed at the origin of the
transverse lattice, as defined in eq.~(\ref{eq:Cpdef}).

\subsection{Definition and basic properties}
% Therefore, 
We draw randomly (within the statistical ensemble defined
by eqs.~(\ref{eq:MVsources})) $N$ configurations $\rho_{i,a}$
($1\le i\le N$) of the source, and the exact analytical formula
(\ref{eq:Cx}) and its Fourier transform are replaced by
\begin{eqnarray}
  &&
C_{_N}(\x_\perp,0)\equiv\frac{1}{N}\sum_{i=1}^N 
{\frac{1}{N_c}\;
{\rm tr}\,\left(U_{i}(\x_\perp)U^\dagger_{i}(0_\perp)\right)}
\nonumber\\
&&C_{_N}(\p_\perp)\equiv\int d^2\x_\perp\;e^{i\p_\perp\cdot\x_\perp}\;C_{_N}(\x_\perp,0)
\; ,
\label{eq:Cx-MC1}
\end{eqnarray}
where $U_i(\x_\perp)$ is a Wilson line calculated with the source
$\rho_{i,a}$. (We use continuous notations for the Fourier transform
for simplicity, but all the plots shown in the paper use a lattice
discretization and a discrete Fourier transform, as explained in the
previous section.) In the following, we will call $C_{_N}(\p_\perp)$ a
\emph{measurement} of $C(\p_\perp)$ with statistics $N$.

$C_{_N}(\p_\perp)$ is itself a random quantity, since it is obtained
from a finite number $N$ of samples of the random source
$\rho_a$. However, if we repeat many times the measurement defined
in eq.~(\ref{eq:Cx-MC1}), the mean value of the quantity
$C_{_N}(\p_\perp)$ is the expected correlator $C(\p_\perp)$:
\begin{equation}
\left<C_{_N}(\p_\perp)\right>=C(\p_\perp).
\end{equation}
In this equation and in the following, the angle brackets
$\langle\cdots\rangle$ indicate a statistical average over repeated
measurements of the quantity contained between the brackets. This
indicates that $C_{_N}(\p_\perp)$ fluctuates around the exact result
$C(\p_\perp)$.  Furthermore, when $N\to\infty$, these fluctuations
should decrease thanks to the property
\begin{equation}
\lim_{N\to \infty} C_{_N}(\p_\perp)=C(\p_\perp).
\end{equation}
In other words, a single measurement with infinitely large statistics
should also yield the exact answer.

In the previous section, we have mentioned the fact that exact
correlation function $C(\p_\perp)$ is real valued thanks to the
symmetry of the ensemble of color sources under
$\x_\perp\to -\x_\perp$. However, this symmetry is not true event by
event, and therefore measurements $C_{_N}(\p_\perp)$ with finite
statistics are complex valued. In the limit of a large number of
samples, $N\to \infty$, we have the following results,
\begin{align}
\lim_{N\to\infty} C_{_N}(\p_\perp)&=C(\p_\perp),\nonumber\\
\lim_{N\to\infty} {\rm Re}\,C_{_N}(\p_\perp)&=C(\p_\perp),\nonumber\\
\lim_{N\to\infty} {\rm Im}\,C_{_N}(\p_\perp)&=0,\nonumber\\
\lim_{N\to\infty} \big|C_{_N}(\p_\perp)\big|&=C(\p_\perp).
\end{align}
Therefore, we may use any of $C_{_N}$, ${\rm Re}\,C_{_N}$ or
$\big|C_{_N}\big|$ as Monte-Carlo approximations of the exact result
$C$. If we insist on a real-valued approximation, then
${\rm Re}\,C_{_N}$ or $\big|C_{_N}\big|$ should be
considered.

Moreover, the ensemble of sources defined by eqs.~(\ref{eq:MVsources})
is invariant under rotations in the transverse plane, implying that
the exact correlation function $C(\p_\perp)$ in fact depends only on
the norm of the transverse momentum $p_\perp\equiv
|\p_\perp|$. Again, this is a symmetry which is not realized
event-by-event and therefore the above Monte-Carlo measurements are
not rotationally invariant. We may enforce a rotationally invariant
result by performing in addition an average over the orientation of
$\p_\perp$ in the transverse plane, by using one of the following
definitions
\begin{align}
  \big({\rm Re}\,C_{_N}(\p_\perp)\big)_\theta&\equiv\int\frac{d\theta}{2\pi}\; {\rm Re}\,C_{_N}(\p_\perp),\nonumber\\
\big(\big|C_{_N}(\p_\perp)\big|\big)_\theta&\equiv\int\frac{d\theta}{2\pi}\; \big|C_{_N}(\p_\perp)\big|,\nonumber\\
  \big(\big|C_{_N}(\p_\perp)\big|^2\big)_\theta^{1/2}&\equiv\Big[\int\frac{d\theta}{2\pi}\; \big|C_{_N}(\p_\perp)\big|^2\Big]^{1/2}.
 \label{eq:ang-avgs}
\end{align}
(Here also, we use a continuous notation for the integration over
$\theta$, but in the lattice implementation this average is in fact a
sum over the finite set of momenta that have a common norm
(\ref{eq:lattice-norm}).) When considering the angular average of the
modulus, two definitions are possible, corresponding to the second or
third of eqs.~(\ref{eq:ang-avgs}). In the rest of this paper, we are
using the third equation, but we have checked that both definitions
behave very similarly. Note also that this angular averaging
eliminates the imaginary part of $C_{_N}(\p_\perp)$. Indeed, this
average restores the symmetry $\x_\perp \to -\x_\perp$ even for a
finite number of samples, leading to
\begin{align}
  \big({\rm Im}\,C_{_N}(\p_\perp)\big)_\theta=0,\quad
  \big({\rm Re}\,C_{_N}(\p_\perp)\big)_\theta=\big(C_{_N}(\p_\perp)\big)_\theta.
\end{align}

\begin{figure}[htbp]
  \begin{center}
    \resizebox*{9cm}{!}{\includegraphics{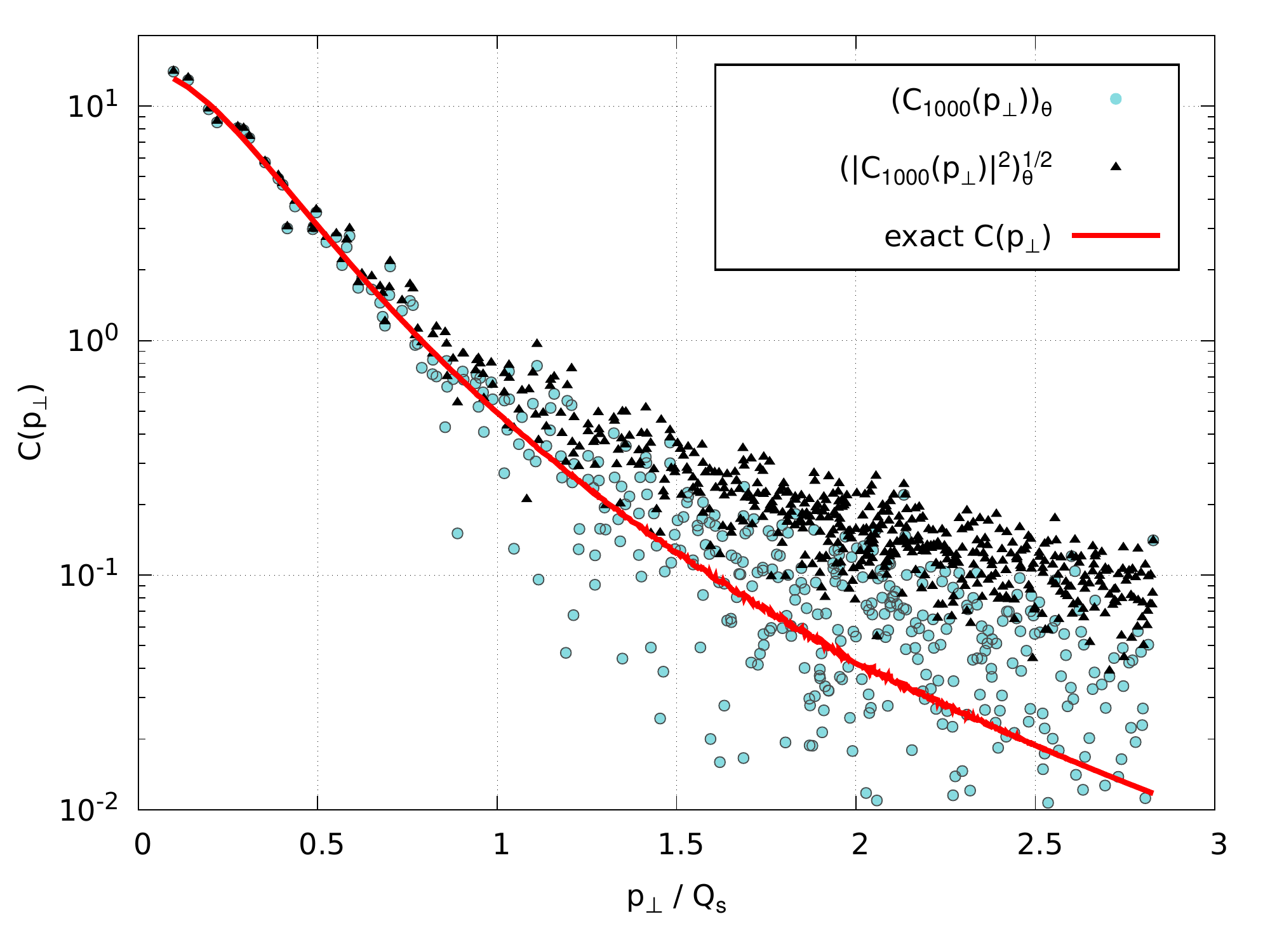}}
  \end{center}
  \caption{\label{fig:Cp1000raw} Angular averaged values of a
    measurement $C_{_N}(\p_\perp)$ for $N=10^3$ (circles), on a
    $64\times 64$ lattice. Triangles: values of
    $\big(\big|C_{_N}(\p_\perp)\big|^2\big)_\theta^{1/2}$, for the
    same number of samples. Solid line: exact value of $C(\p_\perp)$
    on the same lattice. }
\end{figure}

In the figure \ref{fig:Cp1000raw}, we show the angular averaged values
of $C_{_N}(\p_\perp)$, as well as the third of
eqs.~(\ref{eq:ang-avgs}), with $N=10^3$. Although this is not shown in
the plot in order to reduce clutter, we have checked that the angular
averaging has only a mild effect in reducing the dispersion of
the displayed quantities (expect of course for the imaginary part of
$C_{_N}$, which is totally canceled by the averaging).
The first observation one may draw from this plot is the large scatter
of the values of $\big(C_{_N}(\p_\perp)\big)_\theta$, despite the
large number of samples used in the measurement. Another striking
aspect is that the dispersion of the points increases
relatively to the magnitude of the exact result as the momentum
increases. In fact, at large momentum, this dispersion is so large
that even the order of magnitude of the Monte-Carlo estimate is poorly
controlled.

Consider now the triangles, that show the values of
$\big(\big|C_{_N}(\p_\perp)\big|^2\big)_\theta^{1/2}$. At first sight,
they appear to have a significantly lower dispersion compared to
$\big(C_{_N}(\p_\perp)\big)_\theta$. If we did not know the exact
answer, this might mislead us into thinking that
$\big(\big|C_{_N}(\p_\perp)\big|^2\big)_\theta^{1/2}$ ought to provide
a better approximation. However, since in this example we also know
the exact result, the comparison quickly dissipates this hope: indeed,
the plot readily shows that
$\big(\big|C_{_N}(\p_\perp)\big|^2\big)_\theta^{1/2}$ fluctuates
around a mean value that differ significantly from the exact result
(in fact, it even has a tail that behaves as a different power of
momentum).

In the rest of this section, we explain these observations by studying
the statistical distribution of the values of $C_{_N}(\p_\perp)$ and
$\big|C_{_N}(\p_\perp)\big|^2$.

\subsection{Statistical fluctuations of $C_{_N}(\p_\perp)$}
In order to estimate the statistical error made when we replace
$C(\p_\perp)$ by a single measurement $C_{_N}(\p_\perp)$, let us
consider the following variance
\begin{equation}
  \sigma_{_N}^2(\p_\perp)
  \equiv
  \big<\big|C_{_N}(\p_\perp)-C(\p_\perp)\big|^2\big>
  =\big<\big|C_{_N}(\p_\perp)\big|^2\big>-C^2(\p_\perp).
  \label{eq:s2n-def}
\end{equation}
(We use the squared modulus $(C_{_N}-C)(C_{_N}-C)^*$ in order to have
a real positive definite result despite the fact that $C_{_N}$ is
complex-valued.) This quantity can also be written as
\begin{equation}
  \sigma_{_N}^2(\p_\perp)
  =
  \big<\big({\rm Re}\,C_{_N}(\p_\perp)\big)^2\big>-\big<{\rm Re}\,C_{_N}(\p_\perp)\big>^2
  +
   \big<\big({\rm Im}\,C_{_N}(\p_\perp)\big)^2\big>-{\underbrace{\big<{\rm Im}\,C_{_N}(\p_\perp)\big>}_{0}}^2.
\end{equation}
In other words, this quantity is the sum of the variances of the real
part and of the imaginary part of $C_{_N}(\p_\perp)$. Note also that
we have not included an angular average in the definition of this
variance, for simplicity (the quantity defined in
eq.~(\ref{eq:s2n-def}) can be evaluated analytically). It nevertheless
provides a good estimate of the fluctuations of
$\big(C_{_N}(\p_\perp)\big)_\theta$, as we shall see shortly.

Let us first rewrite this variance in terms of correlators in
coordinate space,
\begin{eqnarray}
  \sigma_{_N}^2(\p_\perp)
  =
  \int d^2\x_\perp d^2\y_\perp\; e^{i\p_\perp\cdot(\x_\perp-\y_\perp)}\;
  \Big[\big<C_{_N}(\x_\perp)C_{_N}^*(\y_\perp)\big>
  -\big<C_{_N}(\x_\perp)\big>\big<C_{_N}^*(\y_\perp)\big>\Big].
\end{eqnarray}
In order to evaluate the ensemble average of
$C_{_N}(\x_\perp)C_{_N}^*(\y_\perp)$, we first need to generalize the
third of eqs.~(\ref{eq:MVsources}) into
\begin{eqnarray}  \big<\rho_{i,a}(x^-,\x_\perp)\rho_{j,b}(y^-,\y_\perp)\big>
  = \mu^2 (x^-)\,\delta_{ij}\,\delta_{ab}\delta(x^--y^-)\delta(\x_\perp-\y_\perp).
  \label{eq:MVsources1}
\end{eqnarray}
In other words, two configurations $i$ and $j$ of the source are not
correlated if $i\not= j$, and if $i=j$ they are correlated as per the
usual CGC prescription. Then, we have
\begin{align}
  \left<C_{_N}(\x_\perp)C_{_N}^*(\y_\perp)\right>
  &=\frac{1}{N^2}\sum_{i,j}
      \frac{1}{N_c^2}
      \big<{\rm tr}\,(U_i(\x_\perp)U^\dagger_i(0_\perp))\;{\rm tr}\,(U^\dagger_j(\y_\perp)U_j(0_\perp))\big>\nonumber\\
  &=
  \frac{1}{N^2}\!\!\sum_{i\not= j}
      \frac{1}{N_c^2}
      \big<{\rm tr}\,(U_i(\x_\perp)U^\dagger_i(0_\perp))\big>
  \big<{\rm tr}\,(U^\dagger_j(\y_\perp)U_j(0_\perp))\big>\nonumber\\
  &
  +
     \frac{1}{N^2}\sum_{i}
     \frac{1}{N_c^2}
     \big<{\rm tr}\,(U_i(\x_\perp)U^\dagger_i(0_\perp))\;{\rm tr}\,(U^\dagger_i(\y_\perp)U_i(0_\perp))\big>\nonumber\\
  &=\left(1-\frac{1}{N}\right)\,C(\x_\perp)\,C(\y_\perp)
  +\frac{1}{N}\,{\bs\Sigma}_4(\x_\perp,\y_\perp),
  \label{eq:S4-0}
\end{align}
where we have defined
\begin{equation}
  {\bs\Sigma}_4(\x_\perp,\y_\perp)
  \equiv
  \frac{1}{N_c^2}\big<{\rm tr}\,(U(\x_\perp)U^\dagger(0_\perp))\;{\rm tr}\,(U^\dagger(\y_\perp)U(0_\perp))\big>.
  \label{eq:S4}
\end{equation}
The variance $\sigma_{_N}^2(\p_\perp)$ thus reads
\begin{equation}
\sigma_{_N}^2(\p_\perp)
  =\frac{1}{N}
  \int d^2\x_\perp d^2\y_\perp\; e^{i\p_\perp\cdot(\x_\perp-\y_\perp)}\;
  \Big[{\bs\Sigma}_4(\x_\perp,\y_\perp)-C(\x_\perp)C(\y_\perp)\Big],
  \label{eq:s2n-alt}
\end{equation}
and unsurprisingly it decreases as $1/N$ when the number of samples increases.

A general method for evaluating this type of correlation function in
the MV model can be found in the appendix A of ref.~\cite{Blaizot:2004wv}.
When applied to eq.~(\ref{eq:S4}), this method leads to the following
expression\footnote{With these notations, the 2-point function of eq.~(\ref{eq:Cx})
  can be written as $C(\x_\perp,0)=\exp(C_{_F} \ell_{x0})$.}
\begin{equation}
  {\bs\Sigma}_4(\x_\perp,\y_\perp)
  =\left(1-\frac{1}{N_c^2}\right)\;e^{\lambda_+}
  + \frac{1}{N_c^2}\;e^{\lambda_-},
  \label{eq:S4-1}
\end{equation}
with
\begin{eqnarray}
  &&
  \lambda_+\equiv\frac{N_c\,(\ell_{x0}+\ell_{0y})}{2}-\frac{\ell_{xy}}{2N_c}
  ,\quad\lambda_-\equiv C_{_F}\,\ell_{xy},
  \nonumber\\
  &&\mbox{and}\quad \ell_{xy}\equiv Q_s^2 \int\frac{d^2\p_\perp}{(2\pi)^2}\;\frac{e^{i\p_\perp\cdot(\x_\perp-\y_\perp)}-1}{p_\perp^4}.
    \label{eq:S4-2}
\end{eqnarray}
Note that these formulas are considerably simpler than those for a
completely general 4-point correlator of Wilson lines (see the
subsection \ref{sec:self-var} for the general case), thanks to the
fact that two of the points in eq.~(\ref{eq:S4}) coincide.

\subsubsection{Large $N_c$ limit} 
In the limit of a large number of colors, we
can neglect all the terms that are suppressed by inverse powers of
$N_c$, which leads to
\begin{eqnarray}
&&{\bs\Sigma}_4(\x_\perp,\y_\perp)
  \empile{\approx}\over{N_c\gg 1}e^{\lambda_+}\nonumber\\
  &&\lambda_+\empile{\approx}\over{N_c\gg 1} \frac{N_c\,(\ell_{x0}+\ell_{0y})}{2}
  \; .
\end{eqnarray}
In this limit, ${\bs\Sigma}_4(\x_\perp,\y_\perp)$ factorizes into the
product of a function of $\x_\perp$ and a function of $\y_\perp$,
\begin{equation}
{\bs\Sigma}_4(\x_\perp,\y_\perp)
  \empile{\approx}\over{N_c\gg 1} C(\x_\perp)C(\y_\perp)\; ,
\end{equation}
and the variance $\sigma_{_N}(\p_\perp)$ vanishes in the large $N_c$
limit. In other words, in the limit of a large number of colors, a
single configuration of the source $\rho_a$ is sufficient in order to
obtain the correct average over the sources. This is not surprising
since there are no correlations between different colors in the MV
ensemble: generating a single $\rho_a$ with many color components has
the same effect as generating many configurations at finite $N_c$.

\subsubsection{Finite $N_c$} 
\begin{figure}[htbp]
  \begin{center}
    \resizebox*{9cm}{!}{\includegraphics{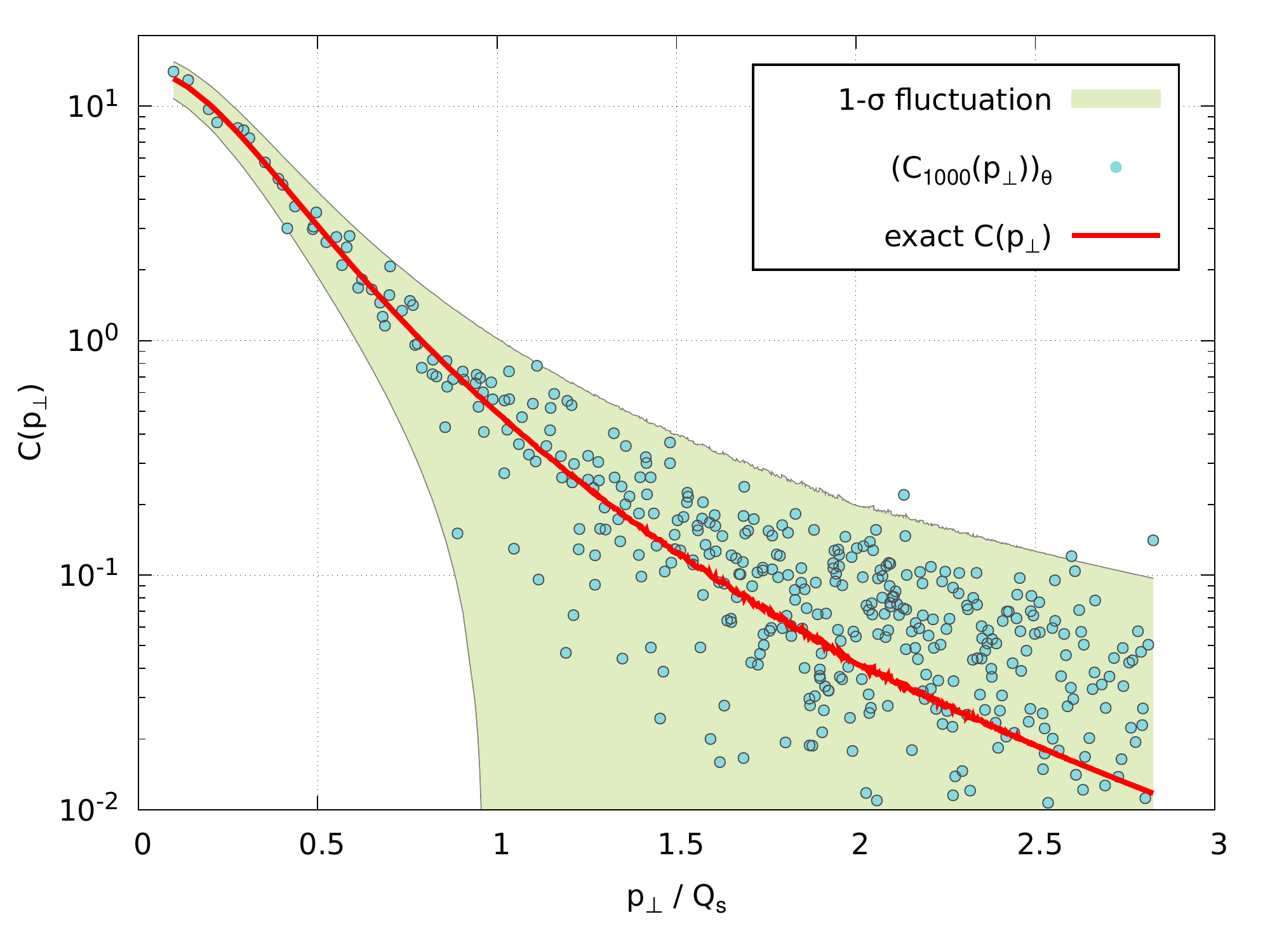}}
  \end{center}
  \caption{\label{fig:Ck1000}Monte-Carlo evaluation of the function
    $C(p_\perp)$ from $\big(C_{_N}(\p_\perp)\big)_\theta$. Shaded
    band: one-sigma statistical fluctuation of $C_{_N}(\p_\perp)$
    obtained form eqs.~(\ref{eq:s2n-alt}), (\ref{eq:S4-1}) and
    (\ref{eq:S4-2}). Solid line: exact result based on
    eq.~(\ref{eq:Cx}). Circles: result of a single measurement of
    $\big(C_{_N}(\p_\perp)\big)_{\theta}$ with $N=10^3$ configurations.}
\end{figure}

The second term of eq.~(\ref{eq:S4-1}) is
particularly interesting because it couples in a simple way the
points $\x_\perp$ and $\y_\perp$. After Fourier transform, this term
gives the following contribution to $\sigma_{_N}^2(\p_\perp)$,
\begin{equation}
  \frac{1}{N\,N_c^2} \;{\cal S}_\perp\;C(\p_\perp),
  \label{eq:s2n-anom}
\end{equation}
where ${\cal S}_\perp$ is the transverse area of the system under
consideration (in order to see this, one should rewrite the integrals
over $\x_\perp,\y_\perp$ as integrals over the difference
$\x_\perp-\y_\perp$ and barycenter $(\x_\perp+\y_\perp)/2$ -- it is
the latter that gives the factor ${\cal S}_\perp$). A crucial property
of this term is that its behavior at large $\p_\perp$
($\sim p_\perp^{-4}$) differs from that of the squared exact result
$C^2(\p_\perp)\sim p_\perp^{-8}$. Therefore, the statistical errors
are comparatively very large in the tail of the function
$C(\p_\perp)$, to a point that makes this Monte-Carlo evaluation
worthless with any reasonable number of samples.  To illustrate this
discussion, we show again in the figure \ref{fig:Ck1000} the estimate
of $C(\p_\perp)$ from one measurement
$\big(C_{_N}(\p_\perp)\big)_\theta$ with $N=10^3$, superimposed over a
band whose boundaries are $C(\p_\perp)\pm \sigma_{_N}(\p_\perp)$ (with
$\sigma_{_N}(\p_\perp)$ evaluated numerically from
eqs.~(\ref{eq:s2n-alt}), (\ref{eq:S4-1}) and (\ref{eq:S4-2})).

\subsection{Statistical fluctuations of  $\big(|C_{_N}|^2\big)_\theta^{1/2}$}
\begin{figure}[htbp]
  \begin{center}
    \resizebox*{9cm}{!}{\includegraphics{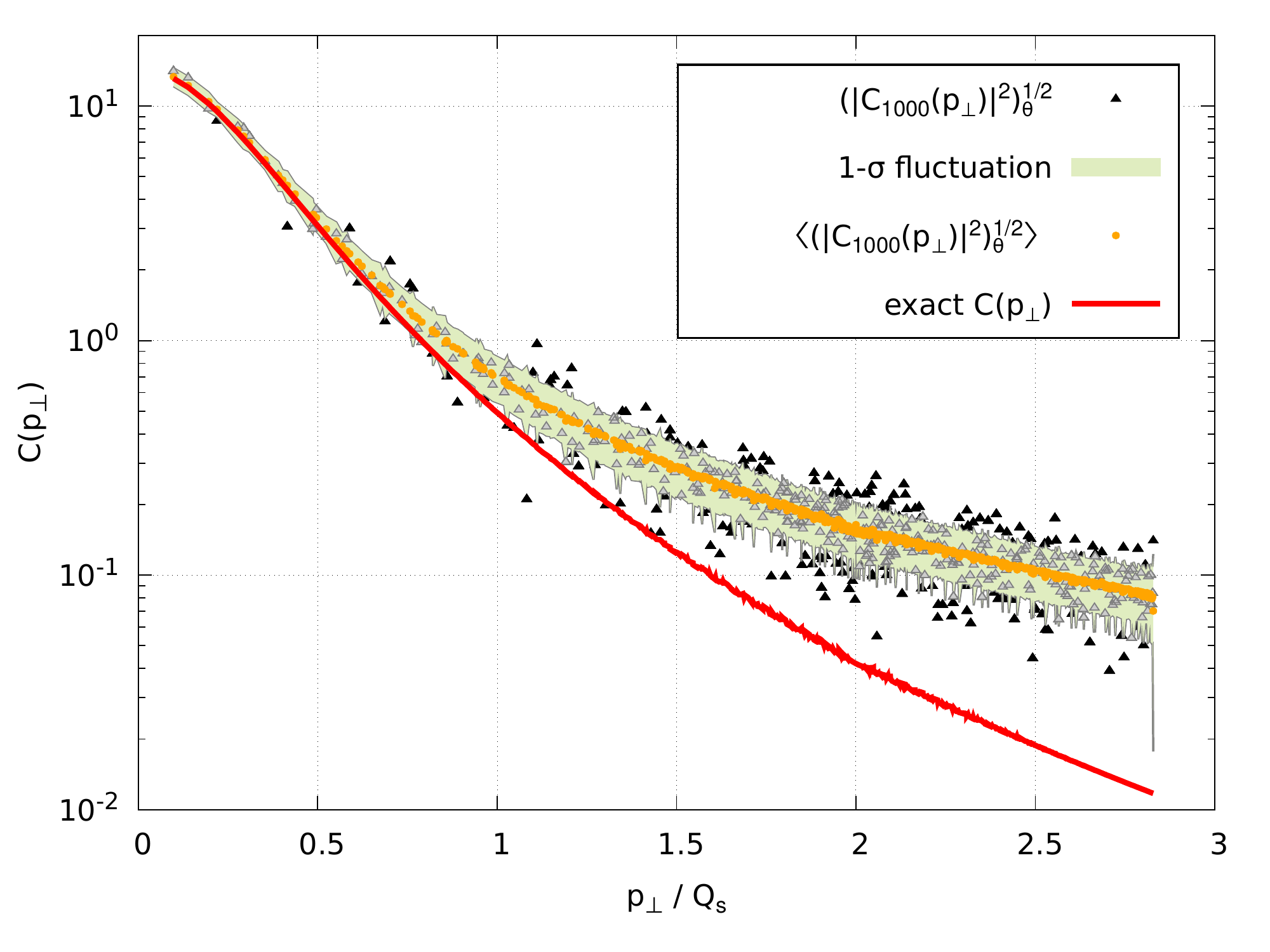}}
  \end{center}
  \caption{\label{fig:Ck1000mod}Monte-Carlo evaluation of the function
    $C(p_\perp)$ from $\big(|C_{_N}|^2\big)_\theta^{1/2}$. Shaded
    band: Monte-Carlo estimate of the one-sigma fluctuation of
    $\big(|C_{_N}|^2\big)_\theta^{1/2}$. Solid line: exact result
    based on eq.~(\ref{eq:Cx}). Triangles: result of a single
    measurement of $\big(|C_{_N}|^2\big)_\theta^{1/2}$ with $N=10^3$
    configurations. Orange dots: mean value of these measurements.}
\end{figure}

In the figure \ref{fig:Ck1000mod}, we show the result of
%$\big|C_{_N}(\p_\perp)\big|$, for a single measurement with
a single measurement of $\big(|C_{_N}|^2\big)_\theta^{1/2}$ with
$N=10^3$ configurations of the sources $\rho_a$. As
already mentioned, this quantity appears much less noisy than the
estimate based on $\big(C_{_N}(\p_\perp)\big)_\theta$ shown in the
previous figures. This can be understood from the variance of
$\big(|C_{_N}|^2\big)_\theta^{1/2}$
\begin{equation}
  \big<\big(|C_{_N}|^2\big)_\theta\big>-\big<\big(|C_{_N}|^2\big)_\theta^{1/2}\big>^2,
  \label{eq:varCN2}
\end{equation}
that we have indicated in the figure \ref{fig:Ck1000mod} by a shaded
area. Here, we have estimated eq.~(\ref{eq:varCN2}) by a Monte-Carlo
sampling, i.e. by performing $M\gg 1$ (in the figure, we have used
$M=2000$) successive measurements of the quantity
$\big(|C_{_N}|^2\big)_\theta^{1/2}$. These $M$ measurements have also
been used to compute the mean value of
$\big(|C_{_N}|^2\big)_\theta^{1/2}$, shown by the orange points in the
figure.

This variance clearly confirms that
$\big(|C_{_N}|^2\big)_\theta^{1/2}$ provides an estimate of the
2-point correlator that fluctuates much less than those based on
$\big(C_{_N}\big)_\theta$ itself. However, the figure also shows
clearly that the mean value of $\big(|C_{_N}|^2\big)_\theta^{1/2}$
differs substantially from the exact value, and in particular exhibits
a different power law of momentum in the tail. This can be understood
from eqs.~(\ref{eq:S4-0}), (\ref{eq:S4-1}), (\ref{eq:S4-2}) and
(\ref{eq:s2n-anom}).
Indeed, these equations indicate that the mean
value of $\big|C_{_N}(\p_\perp)\big|^2$ (this is not exactly the same
quantity as the $\big(|C_{_N}|^2\big)_\theta$ considered in this
subsection, but their tails have the same asymptotic behavior)
contains a term in $N^{-1}p_\perp^{-4}$, while the exact answer for
$C^2(\p_\perp)$ has a tail in $p_\perp^{-8}$.  The
measurement of $\big(|C_{_N}|^2\big)_\theta^{1/2}$ contains a
contamination at large momentum that goes away rather slowly (like
$N^{-1}$) with the number $N$ of source configurations used in each
measurement.  In other words, the comparatively small fluctuation of
$\big(|C_{_N}|^2\big)_\theta^{1/2}$ is rather misleading, because it
is not an indicator of its proximity with the correct value
$C(\p_\perp)$.

\section{Improvement by averaging over the barycenter}
\label{sec:self}
\subsection{Definition}
In the definition (\ref{eq:Cpdef}) of the 2-point function that we
have used as example, one of the two points is fixed at the origin of
the transverse plane. In this section, we shall discuss the
improvement achieved by letting this point free and integrating it out, i.e. by
generalizing the definition of the correlation function as follows
\begin{equation}
  C(\p_\perp)\equiv\frac{1}{N_c}\frac{1}{{\cal S}_\perp}
  \int d^2\x_\perp d^2\y_\perp\;e^{i\p_\perp\cdot(\x_\perp-\y_\perp)}\;
  \left<{\rm tr}\,\left(U(\x_\perp)U^\dagger(\y_\perp)\right)\right>.
  \label{eq:Cpdef1}
\end{equation}
%where ${\cal S}_\perp$ is the transverse area of the system. 
The two definitions are equivalent for a system invariant by
translation\footnote{On the lattice, they are equivalent
  provided that one uses periodic boundary conditions.}. However, in a
Monte-Carlo evaluation, they may differ with finite statistics since
individual configurations of the sources are not translation
invariant.

Interestingly, this averaging over the barycenter eliminates the
imaginary part of Monte-Carlo estimates, even when the correlator is
evaluated with a finite number of samples. Indeed, the complex
conjugate of $C_{_N}(\p_\perp)$ reads
\begin{align}
  C_{_N}^*(\p_\perp)&=\frac{1}{NN_c{\cal S}_\perp}\sum_{i}
  \int d^2\x_\perp d^2\y_\perp\;e^{-i\p_\perp\cdot(\x_\perp-\y_\perp)}\;
       {\rm tr}\,\left((U_i(\x_\perp)U^\dagger_i(\y_\perp))^\dagger\right)
       \nonumber\\
       &=\frac{1}{NN_c{\cal S}_\perp}\sum_{i}
  \int d^2\x_\perp d^2\y_\perp\;e^{-i\p_\perp\cdot(\x_\perp-\y_\perp)}\;
       {\rm tr}\,\left(U_i(\y_\perp)U^\dagger_i(\x_\perp)\right)
       \nonumber\\
       &=\frac{1}{NN_c{\cal S}_\perp}\sum_{i}
  \int d^2\x_\perp d^2\y_\perp\;e^{i\p_\perp\cdot(\x_\perp-\y_\perp)}\;
       {\rm tr}\,\left(U_i(\x_\perp)U^\dagger_i(\y_\perp)\right)
       \nonumber\\
       &= C_{_N}(\p_\perp)\; .
\end{align}
Note that the manipulations performed here are only possible because
we are integrating on both $\x_\perp$ and $\y_\perp$.

\begin{figure}[htbp]
  \begin{center}
    \resizebox*{9cm}{!}{\includegraphics{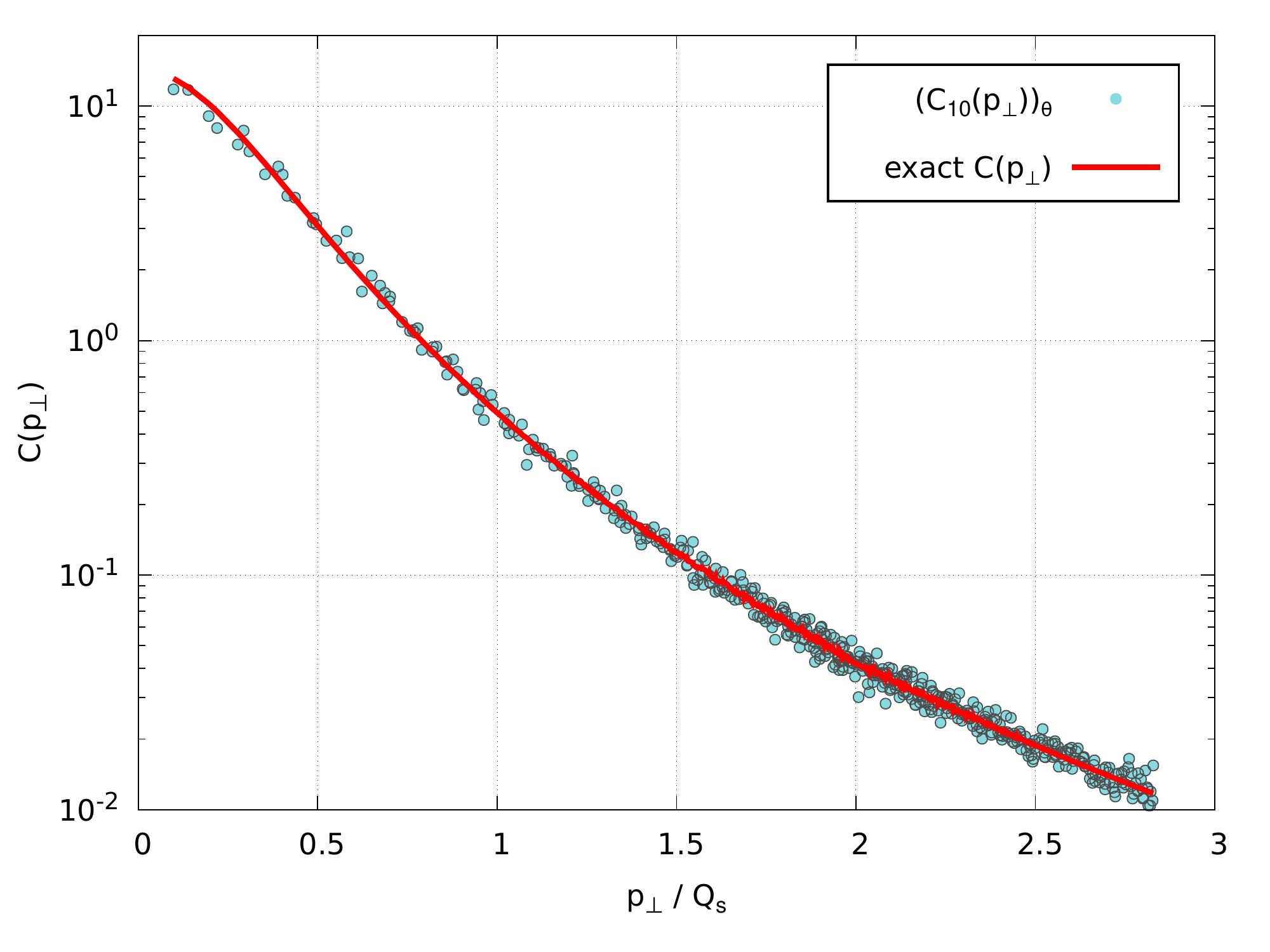}}
  \end{center}
  \caption{\label{fig:Ck1xy}Monte-Carlo evaluation of the function
    $C(p_\perp)$ from $\big(C_{_N}(\p_\perp)\big)_\theta$, with barycenter
    and angular averaging. Solid line: exact result based on
    eq.~(\ref{eq:Cx}). Circles: result of a single measurement of
    $C_{_N}(\p_\perp)$ with only ten ($N=10$)
    configurations.}
\end{figure}
In the figure \ref{fig:Ck1xy}, we show the estimate of $C(\p_\perp)$
from the  value of $\big(C_{_N}(\p_\perp)\big)_\theta$, for $N=10$, using the
barycenter averaging described in this section. One can see that
despite the small number of configurations used in this computation,
this estimate tracks remarkably well the exact result.  Naturally, one
naively expects that the barycenter averaging roughly amounts to increasing
the statistics by a factor equal to the number of lattice sites,
i.e. by a factor $64\times 64=4096$ in the present case. But the
figure also suggests that the momentum dependence of the fluctuations
has changed, in such a way that the relative error is now independent
of momentum.

\subsection{Statistical fluctuations}
\label{sec:self-var}
\subsubsection{Analytical expression}
The momentum dependence of the fluctuations in measurements using
barycenter averaging can be understood by calculating the
corresponding variance. With barycenter averaging, the mean value of
$C_{_N}^2(\p_\perp)$ reads
\begin{align}
  \big<C_{_N}^2(\p_\perp)\big>
  =
  \left(1-\frac{1}{N}\right)\,C^2(\p_\perp)
  +\frac{1}{N}\,{\bs\Sigma}_4(\p_\perp),
  \label{eq:var-xy}
\end{align}
where we have defined
\begin{align}
  {\bs\Sigma}_4(\p_\perp)
  &\equiv
  \frac{1}{{\cal S}_\perp^2}
  \int d^2\x_\perp d^2\y_\perp d^2\u_\perp d^2\v_\perp\;
  e^{i\p_\perp\cdot(\x_\perp-\y_\perp)}e^{i\p_\perp\cdot(\u_\perp-\v_\perp)}
  \nonumber\\
  &\qquad\qquad\qquad
  \times
  \underbrace{\tfrac{1}{N_c^2}\left<{\rm tr}\,\big(U(\x_\perp)U^\dagger(\y_\perp)\big)
        \;{\rm tr}\,\big(U(\u_\perp)U^\dagger(\v_\perp)\big)\right>}_{{\bs\Sigma}_4(\x_\perp\y_\perp;\u_\perp\v_\perp)}.
\end{align}
Using the method described in the appendix A of
ref.~\cite{Blaizot:2004wv}, we obtain
\begin{equation}
  {\bs\Sigma}_4(\x_\perp\y_\perp;\u_\perp\v_\perp)
  =
  A_+\;e^{ \lambda_+}
  +
  A_-\;e^{ \lambda_-}\; ,
\end{equation}
with
\begin{align}
  A_\pm &\equiv
  \frac{1}{2}\pm \frac{1}{\sqrt{\Delta}}
  \left(\frac{N_c(\alpha-\beta)}{2}+\frac{\beta-\gamma}{N_c}\right)
  \nonumber\\
  \lambda_\pm&\equiv \frac{N_c(\alpha+\beta)}{4}+\frac{\gamma-\alpha-\beta}{2N_c}\pm \frac{\sqrt{\Delta}}{4}\nonumber\\
  \Delta&\equiv N_c^2(\alpha-\beta)^2+4(\alpha-\gamma)(\beta-\gamma)
  \nonumber\\
  \alpha&\equiv\ell_{xy}+\ell_{uv}\;,\quad
  \beta\equiv\ell_{xv}+\ell_{yu}\;,\quad
  \gamma\equiv\ell_{xu}+\ell_{yv},
\end{align}
and $\ell_{xy}$ is the quantity defined in eq.~(\ref{eq:S4-2}).

\subsubsection{Large $N_c$ limit}
In the large $N_c$ limit, we have
\begin{eqnarray}
  &&
  \lambda_+\approx\frac{N_c}{2}\,{\rm max}\,(\alpha,\beta)\; ,\quad
  \lambda_-\approx\frac{N_c}{2}\,{\rm min}\,(\alpha,\beta)
  \nonumber\\
  && A_\pm \approx \frac{1}{2}(1\pm{\rm sign}\,(\alpha-\beta)),
\end{eqnarray}
and as a consequence we obtain
\begin{equation}
  {\bs\Sigma}_4(\x_\perp\y_\perp;\u_\perp\v_\perp)
  \approx
  e^{\tfrac{ Q_s^2  N_c}{2}(\ell_{xy}+\ell_{uv})}\; .
\end{equation}
After performing the Fourier transforms and inserting into
eq.~(\ref{eq:var-xy}), this leads to a vanishing variance
\begin{equation}
\big<C_{_N}^2(\p_\perp)\big>
  - C^2(\p_\perp)\approx 0\; .
\end{equation}
In other words, the fluctuations of $C_{_N}$ are suppressed by inverse
powers of the number of colors, as was already the case when we did
not perform any barycenter averaging.

\subsubsection{Finite $N_c$}
We thus need to keep $N_c$ finite in order to obtain a non trivial
result for the variance. If we expand to order $N_c^{-2}$ the
coefficients $A_\pm$, while keeping only the leading $N_c$ term in the
eigenvalues $\lambda_\pm$, we obtain
\begin{align}
  &{\bs\Sigma}_4(\x_\perp\y_\perp;\u_\perp\v_\perp)
  \approx
  e^{\tfrac{ Q_s^2  N_c}{2}(\ell_{xy}+\ell_{uv})}
  \nonumber\\
  &\quad
  +\frac{1}{N_c^2}
  \left(\frac{\ell_{xv}+\ell_{yu}-\ell_{xu}-\ell_{yv}}{\ell_{xv}+\ell_{yu}-\ell_{xy}-\ell_{uv}}\right)^2
  \left(
  e^{\tfrac{N_c}{2}(\ell_{xv}+\ell_{yu})}
  -
  e^{\tfrac{N_c}{2}(\ell_{xy}+\ell_{uv})}\right),
\end{align}
so that the variance can be approximated by
\begin{align}
  &
  \!\!\big<C_{_N}^2(\p_\perp)\big>-C^2(\p_\perp)
  \approx\frac{1}{NN_c^2 }
  \int \frac{d^2\x_\perp d^2\y_\perp d^2\u_\perp d^2\v_\perp}{{\cal S}_\perp^2}\;
  e^{i\p_\perp\cdot(\x_\perp-\y_\perp+\u_\perp-\v_\perp)}
  \nonumber\\
  &\qquad\quad\times
  \left(\frac{\ell_{xv}+\ell_{yu}-\ell_{xu}-\ell_{yv}}{\ell_{xv}+\ell_{yu}-\ell_{xy}-\ell_{uv}}\right)^2
  \left(e^{\tfrac{N_c}{2}(\ell_{xv}+\ell_{yu})}
  -
  e^{\tfrac{N_c}{2}(\ell_{xy}+\ell_{uv})}\right).
  \label{eq:var-xy-2}
\end{align}
From this formula, one can understand the different behaviors of the
variance with and without barycenter averaging. The integral in this
formula depends on the transverse area ${\cal S}_\perp$, on the
saturation momentum $ Q_s $ and on the transverse momentum $p_\perp$,
combined in such a way to give a result of mass dimension $-4$. The
momentum $\p_\perp$ is the Fourier conjugate of the coordinate
differences $\x_\perp-\y_\perp$ and $\u_\perp-\v_\perp$, or of
$\x_\perp-\v_\perp$ and $\u_\perp-\y_\perp$. In this discussion, one
can use the following approximation for the function $\ell_{xy}$,
\begin{equation}
  \ell_{xy}\approx -\frac{Q_s^2(\x_\perp-\y_\perp)^2}{8\pi}\,
  \ln\left(\frac{a_{_{\rm IR}}}{|\x_\perp-\y_\perp|}\right)\; ,
\end{equation}
where $a_{_{\rm IR}}$ is an infrared cutoff (here introduced in the form
of a distance such that $ Q_s  a_{_{\rm IR}}\gg 1$).  For the sake of this
argument, one can ignore the first factor in the second line of eq.~(\ref{eq:var-xy-2}), because
its dependence on the coordinates is a rational fraction, while the
second factor has an exponential dependence. For momenta $p_\perp
\gtrsim  Q_s $, the Fourier transform of $\exp(N_c \ell_{xy}/2)$
behaves as
\begin{equation}
  \int d^2(\x_\perp-\y_\perp)\;e^{i\p_\perp\cdot(\x_\perp-\y_\perp)}\;
  e^{\tfrac{N_c}{2}\ell_{xy}}\sim \frac{ Q_s^2 }{p_\perp^4}\; .
\end{equation}
Therefore, at large momentum, $ Q_s^2 /p_\perp^4$ is the only
combination by which $ Q_s $ and $p_\perp$ can enter in the
variance. One should therefore count how many of these factors can
arise (1 or 2), and add the appropriate factors of ${\cal S}_\perp$ to
reach the dimension $-4$.  In eq.~(\ref{eq:var-xy-2}), we see that
there are two Fourier integrals with respect to coordinate
differences, each of which brings a factor $ Q_s^2 /p_\perp^4$. The
remaining two integrations are over the ``barycenter''
coordinates. Each of them brings a factor ${\cal S}_\perp$, and they
cancel the factor ${\cal S}_\perp^{-2}$. Therefore, the formula
(\ref{eq:var-xy-2}) behaves as follows
\begin{equation}
  \sqrt{\big<C_{_N}^2(\p_\perp)\big>-C^2(\p_\perp)}
  \sim\frac{1}{N_c\sqrt{N}}\;\frac{ Q_s^2 }{p_\perp^4}.
\label{eq:var-xy-3}
\end{equation}
This explain why in the figure \ref{fig:Ck1xy} the dispersion of the
points appears to be a roughly constant fraction of the central value,
since
\begin{equation}
  \left.\frac{\sqrt{\big<C_{_N}^2(\p_\perp)\big>-C^2(\p_\perp)}}{C(\p_\perp)}\right|_{{\rm with\ barycenter}\atop{\rm averaging}}
  \sim\frac{1}{N_c\sqrt{N}}.
\label{eq:var-xy-4}
\end{equation}

From the same starting point, eq.~(\ref{eq:var-xy-2}), it is easy to
see what changes if we do not use barycenter averaging. In this case, the
coordinates $\y_\perp$ and $\u_\perp$ are not integrated out but
instead fixed to $\y_\perp=\u_\perp=0_\perp$ (the denominator ${\cal
  S}_\perp^2$ also disappears if we do this). Therefore, we have
$\ell_{yu}\equiv 0$, and one of the Fourier integrals disappears in
the first term, which means that we get a single factor
$ Q_s^2 /p_\perp^4$ and a factor ${\cal S}_\perp$ to make up for the
correct dimension. In the end, we now get
\begin{equation}
  \left.
  \frac{\sqrt{\big<C_{_N}^2(\p_\perp)\big>-C^2(\p_\perp)}}{C(\p_\perp)}\right|_{{\rm without\ barycenter}\atop{\rm averaging}}
  \sim\frac{1}{N_c\sqrt{N}}\;\sqrt{\frac{{\cal S}_\perp p_\perp^4}{ Q_s^2 }},
\label{eq:var-xy-5}
\end{equation}
and the variance is increased by a (large) factor
$\sqrt{{\cal S}_\perp}$. The absence of this factor ${\cal S}_\perp$
in eq.~(\ref{eq:var-xy-4}) is the reason why this procedure is called
\emph{barycenter averaging} (or self-averaging): the same
configuration of sources provides an effective statistics enhanced by
the number of lattice sites if we also average over the mid-point in
the 2-point correlation function.  But more importantly, if we do not
perform this barycenter averaging, the momentum dependence of the
variance differs from that of the mean value $C(p_\perp)$ (it has a
large momentum tail that decreases much slower), which leads to very
important relative errors at large momentum.

\section{Summary and conclusions}
\label{sec:concl}
In this paper, we have studied the statistical errors encountered in
the Monte-Carlo evaluation of the expectation value of correlation
functions in the Color Glass Condensate, on the simple example of the
dipole amplitude. Assuming a uniform saturation momentum, this
correlator is invariant by translation in the transverse plane.

In a first series of Monte-Carlo estimates, we do not exploit the
translation invariance and instead we pin one of the two coordinates
to the origin of the transverse plane. A first ``measurement'' we have
considered is simply to compute the correlator with $N$ samples of the
color sources. We observed that this leads to a very noisy result
--even with a rather large value of $N$-- at high momentum, only
marginally improved by averaging over the orientations of the
transverse momentum. More importantly, the relative error appears to
increase dramatically with momentum. This behavior was then explained
by calculating the variance of such measurements.

More unexpected was the behavior of statistical fluctuations in the
case of the modulus of the above Monte-Carlo estimate (still with one
point at a fixed location on the lattice). With the same number $N$ of
samples, it appears to be considerably less noisy, but also rather far
from the exact answer -- and with a relative discrepancy that
increases with momentum. The calculation of the mean value of this
measurement indicates that it indeed differs from the exact answer by
a term of order $N^{-1}$, that dominates the large momentum tail.

Then, we turned to a measurement in which one also averages over the
barycenter of the two points. In this case, the statistical errors are
much lower even with a small number of samples, and in addition they
appear to scale proportionally to the magnitude of the exact result --
unlike the error encountered when one point of the correlator was held
fixed. This improvement due to the averaging over the barycenter could
be explained by studying the variance of these measurements.

Our study was centered on a very simple example, in order to have an
exact answer to compare with and to be able to discuss
semi-analytically the variance of the Monte-Carlo
measurements. However, we expect our observations to be valid for any
translation invariant CGC correlator, namely that the barycenter
averaging reduces the relative statistical error of Monte-Carlo
measurements by a factor $(Q_s^2/{\cal S}_\perp p_\perp^4)^{1/2}$,
where $p_\perp$ denotes generically the Fourier conjugate variable to
a difference of coordinates.  Therefore, our study should be viewed as
a call for caution in the numerical evaluation of correlation
functions in the CGC framework. The first message
is that pinning a point of the correlator at a fixed location
generally leads to significantly larger errors, especially at large
momentum. In most of these measurements, the large statistical errors
will be strikingly visible in the form of a very noisy output, as in
the figure \ref{fig:Ck1000}. But we also would like to warn the reader
about the existence of alternate measurements --also without
barycenter averaging--, where the statistical fluctuations appear to
be much smaller and yet the result is equally incorrect, as in the
figure \ref{fig:Ck1000mod}. In other words, the ``noisiness'' of the
results is not always a good measure of the error. In order to avoid
these problems, the only safe way to limit the statistical errors in
this type of measurement is to average the correlator over the
barycenter of the points it involves.

\section*{Acknowledgements}
FG's work was supported by the Agence Nationale de la Recherche project
ANR-16-CE31-0019-01.

%\bibliographystyle{unsrt}
%\bibliography{biblio}

\begin{thebibliography}{10}

\bibitem{Iancu:2002xk}
Edmond Iancu, Andrei Leonidov, and Larry McLerran.
\newblock {The Color glass condensate: An Introduction}.
\newblock In {\em {QCD perspectives on hot and dense matter. Proceedings, NATO
  Advanced Study Institute, Summer School, Cargese, France, August 6-18,
  2001}}, pages 73--145, 2002.

\bibitem{Weigert:2005us}
Heribert Weigert.
\newblock {Evolution at small x(bj): The Color glass condensate}.
\newblock {\em Prog. Part. Nucl. Phys.}, 55:461--565, 2005.

\bibitem{Gelis:2010nm}
Francois Gelis, Edmond Iancu, Jamal Jalilian-Marian, and Raju Venugopalan.
\newblock {The Color Glass Condensate}.
\newblock {\em Ann. Rev. Nucl. Part. Sci.}, 60:463--489, 2010.

\bibitem{Gelis:2012ri}
F.~Gelis.
\newblock {Color Glass Condensate and Glasma}.
\newblock {\em Int. J. Mod. Phys.}, A28:1330001, 2013.

\bibitem{McLerran:1993ni}
Larry~D. McLerran and Raju Venugopalan.
\newblock {Computing quark and gluon distribution functions for very large
  nuclei}.
\newblock {\em Phys. Rev.}, D49:2233--2241, 1994.

\bibitem{McLerran:1993ka}
Larry~D. McLerran and Raju Venugopalan.
\newblock {Gluon distribution functions for very large nuclei at small
  transverse momentum}.
\newblock {\em Phys. Rev.}, D49:3352--3355, 1994.

\bibitem{JalilianMarian:1997jx}
Jamal Jalilian-Marian, Alex Kovner, Andrei Leonidov, and Heribert Weigert.
\newblock {The BFKL equation from the Wilson renormalization group}.
\newblock {\em Nucl. Phys.}, B504:415--431, 1997.

\bibitem{JalilianMarian:1997gr}
Jamal Jalilian-Marian, Alex Kovner, Andrei Leonidov, and Heribert Weigert.
\newblock {The Wilson renormalization group for low x physics: Towards the high
  density regime}.
\newblock {\em Phys. Rev.}, D59:014014, 1998.

\bibitem{Iancu:2000hn}
Edmond Iancu, Andrei Leonidov, and Larry~D. McLerran.
\newblock {Nonlinear gluon evolution in the color glass condensate. 1.}
\newblock {\em Nucl. Phys.}, A692:583--645, 2001.

\bibitem{Ferreiro:2001qy}
Elena Ferreiro, Edmond Iancu, Andrei Leonidov, and Larry McLerran.
\newblock {Nonlinear gluon evolution in the color glass condensate. 2.}
\newblock {\em Nucl. Phys.}, A703:489--538, 2002.

\bibitem{Krasnitz:2000gz}
Alex Krasnitz and Raju Venugopalan.
\newblock {The Initial gluon multiplicity in heavy ion collisions}.
\newblock {\em Phys. Rev. Lett.}, 86:1717--1720, 2001.

\bibitem{Krasnitz:2001qu}
Alex Krasnitz, Yasushi Nara, and Raju Venugopalan.
\newblock {Coherent gluon production in very high-energy heavy ion collisions}.
\newblock {\em Phys. Rev. Lett.}, 87:192302, 2001.

\bibitem{Lappi:2003bi}
T.~Lappi.
\newblock {Production of gluons in the classical field model for heavy ion
  collisions}.
\newblock {\em Phys. Rev.}, C67:054903, 2003.

\bibitem{Gelis:2001da}
F.~Gelis and A.~Peshier.
\newblock {Probing colored glass via q anti-q photoproduction}.
\newblock {\em Nucl. Phys.}, A697:879--901, 2002.

\bibitem{Gelis:2002ki}
Francois Gelis and Jamal Jalilian-Marian.
\newblock {Photon production in high-energy proton nucleus collisions}.
\newblock {\em Phys. Rev.}, D66:014021, 2002.

\bibitem{Fukushima:2007ki}
Kenji Fukushima.
\newblock {Randomness in infinitesimal extent in the McLerran-Venugopalan
  model}.
\newblock {\em Phys. Rev.}, D77:074005, 2008.

\bibitem{Blaizot:2004wv}
Jean~Paul Blaizot, Francois Gelis, and Raju Venugopalan.
\newblock {High-energy pA collisions in the color glass condensate approach. 2.
  Quark production}.
\newblock {\em Nucl. Phys.}, A743:57--91, 2004.

\end{thebibliography}

\end{document}